\documentclass[twocolumn,english,superscriptaddress, prl, aps]{revtex4}
\usepackage[T1]{fontenc}
\usepackage[latin9]{inputenc}
\usepackage{amsmath}
\usepackage{amssymb}
\usepackage{graphicx}
\usepackage{esint}

\makeatletter
\@ifundefined{textcolor}{}
{%
 \definecolor{BLACK}{gray}{0}
 \definecolor{WHITE}{gray}{1}
 \definecolor{RED}{rgb}{1,0,0}
 \definecolor{GREEN}{rgb}{0,1,0}
 \definecolor{BLUE}{rgb}{0,0,1}
 \definecolor{CYAN}{cmyk}{1,0,0,0}
 \definecolor{MAGENTA}{cmyk}{0,1,0,0}
 \definecolor{YELLOW}{cmyk}{0,0,1,0}
 }

\usepackage{mciteplus}
\usepackage{hyperref}

\makeatother

\usepackage{babel}
\begin{document}

\title{Friedel oscillations due to Fermi arcs in Weyl semimetals}

\author{Pavan Hosur}

\address{Department of Physics, University of California at Berkeley, Berkeley,
CA 94720, USA}

\altaffiliation{Now at Geballe Laboratory for Advanced Materials, Stanford, CA 94305, USA}

\selectlanguage{english}%
\begin{abstract}
Weyl semimetals harbor unusual surface states known as Fermi arcs,
which are essentially disjoint segments of a two dimensional Fermi
surface. We describe a prescription for obtaining Fermi arcs of arbitrary
shape and connectivity by stacking alternate two dimensional electron
and hole Fermi surfaces and adding suitable interlayer coupling. Using
this prescription, we compute the local density of states -- a quantity
directly relevant to scanning tunneling microscopy -- on a Weyl semimetal
surface in the presence of a point scatterer and present results for
a particular model that is expected to apply to pyrochlore iridate
Weyl semimetals. For thin samples, Fermi arcs on opposite surfaces
conspire to allow nested backscattering, resulting in strong Friedel
oscillations on the surface. These oscillations die out as the sample
thickness is increased and Fermi arcs from the bottom surface retreat
and weak oscillations, due to scattering between the top surface Fermi
arcs alone, survive. The surface spectral function -- accessible to
photoemission experiments -- is also computed. In the thermodynamic
limit, this calculation can be done analytically and separate contributions
from the Fermi arcs and the bulk states can be seen.
\end{abstract}
\maketitle
\emph{Weyl semimetals} (WSMs) are rapidly gaining popularity \cite{VolovikBook,PyrochloreWeyl,KrempaWeyl}
as a new, gapless topological phase of matter, as opposed to topological
insulators, which are gapped. A WSM is defined as a phase that has
a pair of non-degenerate bands touching at a certain number of points
in its Brillouin zone. Each such point or Weyl node has a \emph{chirality}
or a \emph{handedness}; very general conditions constrain the right-
and the left-handed Weyl nodes to be equal in number \cite{NielsenFermionDoubling1,NielsenFermionDoubling2}.
Near the nodes, the Hamiltonian resembles that of Weyl fermions well-known
in high-energy physics. These nodes are topologically stable as long
as translational symmetry is conserved, and can only be destroyed
by annihilating them in pairs. Several theoretical proposals for realizing
WSMs now exist in the literature \cite{PyrochloreWeyl,KrempaWeyl,WeylMultiLayer,BernevigDoubleWeyl,CarpentierWeyl,ChoTItoWeyl,HalaszWeyl,JiangWeyl,PhotonicCrystalWeyl}.
WSMs have already been predicted to exhibit several interesting bulk
properties, ranging from unusual quantum hall effects \cite{RanQHWeyl,FangChernSemimetal}
to various effects that rely on a 3D chiral anomaly present in this
phase \cite{NielsenABJ,AjiABJAnomaly,QiWeylAnomaly,RanQHWeyl,SonSpivakWeylAnomaly,ZyuninBurkovWeylTheta,WeylCDW}.
Preliminary bulk transport studies of WSMs have been performed both
theoretically \cite{HosurWeylTransport,BurkovNodalSemimetal,WeylMultiLayer},
as well as experimentally in some candidate materials \cite{WeylResistivityMaeno,EuIridateExperiments}.

A remarkable feature of WSMs is the existence of unconventional surface
states known as \emph{Fermi arcs} (FAs). These FAs are of a different
origin from the FAs that exist in cuprate superconductors. A FA on
a WSM is essentially a segment of a 2D Fermi surface (FS) that connects
the projections of a pair of bulk Weyl nodes of opposite chiralities
onto the surface Brillouin zone \cite{PyrochloreWeyl}. Although FAs
always connect Weyl nodes of opposite chiralities, their exact shapes
and connectivities depend strongly on the local boundary conditions.
Such disconnected segments of zero energy states cannot exist in isolated
2D systems, which must necessarily have a well-defined FS. A WSM in
a slab geometry, however, \emph{is} an isolated 2D system and indeed,
FAs on opposite surfaces together do form a well-defined 2D FS. A
natural question to therefore ask is, {}``what signatures does this
unusual Fermi surface have in scanning tunneling microscopy (STM)
and angle-resolved photoemission spectroscopy (ARPES) -- two common
techniques that can probe surface states directly?''

In this work, we answer this question by computing the local density
of states (LDOS) on the surface of WSM, $\rho_{\mbox{surf}}(\boldsymbol{r},E)$,
in the presence of a point scatterer on the surface as well as the
surface spectral function for a clean system, $A_{\mbox{surf}}^{0}(\boldsymbol{k},E)$.
We apply our results to the iridates, A$_{2}$Ir$_{2}$O$_{7}$, A$=$Y,
Eu, which are predicted to be WSMs \cite{PyrochloreWeyl,KrempaWeyl}.
Both $\rho_{\mbox{surf}}$ and $A_{\mbox{surf}}^{0}$ evolve as the
sample thickness is increased, and the evolution is explained in terms
of the amplitude of the FAs on the far surface diminishing on the
near surface. The calculation is done using a prescription that can
give FAs of arbitrary shape and connectivity and simultaneously generate
the corresponding Weyl nodes in the bulk. The procedure, in a nutshell,
entails stacking electron and hole FSs alternately, and gapping them
out pairwise via interlayer couplings that are designed to leave the
desired FAs on the end layers. The resulting Hamiltonian is of a simple
tight-binding form, which allows us to calculate $A_{\mbox{surf}}^{0}$
analytically in the thermodynamic limit. This quantity
can be directly measured by ARPES.

FAs appear in two qualitatively distinct ways: (a) either FAs on opposite
surfaces overlap, resulting in a gapless semimetal, or, (b) FAs on
opposite surface do not overlap, resulting in a 2D metal with a FS.
The 2D particle density in this metal is proportional to the FS area
according to Luttinger's theorem, and lives predominantly on the surface.
In general, however, some particles will leak into the bulk, but the
bulk filling will typically be $\mathcal{O}(1/L)$, where $L$ is
the slab thickness, and will thus vanish in the thermodynamic limit:
$L\to\infty$. In the model presented here, (a) results when equal
numbers of electron and hole FSs are stacked while (b) is obtained
when the total number of 2D FSs is odd. This is consistent with the
statement made earlier that the FA structure depends strongly on the
boundary conditions, since peeling off a single layer interchanges
(a) and (b).

FAs, in principle, can be generated by: (i) starting with a bulk model
with the desired number of Weyl nodes, (ii) discretizing it in real
space in the finite direction, and (iii) applying suitable boundary
conditions to obtain FAs of the desired structure. While this approach
works in principle, it has several associated complications. For example,
determining the boundary conditions that result in the desired connectivity
of the FAs is non-trivial. For instance, a WSM with four Weyl nodes
$W_{1,2}^{\chi}$ at momenta $\chi\boldsymbol{Q}_{1,2}$, where $\chi=\pm$
denotes the chirality of the Weyl node, has two pairs of FAs on any
surface on which the projections of the Weyl points are distinct.
These FAs can pair up the Weyl points in two qualitatively different
ways: as $(W_{1}^{+}W_{1}^{-})$ and $(W_{2}^{+}W_{2}^{-})$ on each
surface, which is an (a)-type connectivity, or as $(W_{1}^{+}W_{1}^{-})$
and $(W_{2}^{+}W_{2}^{-})$ on the top surface and as $(W_{1}^{+}W_{2}^{-})$
and $(W_{1}^{-}W_{2}^{+})$ on the bottom surface, which falls in
the (b) category. However, there is currently no general prescription
for determining the boundary conditions that give one or the other
connectivity. Moreover, to our knowledge there is also no general
prescription for deriving lattice models with arbitrary numbers and
locations of Weyl nodes. These gaps in working methods are filled
by our top-down approach for generating FAs directly. Our approach
should be useful to model FAs in real systems, where surface effects
can bend the FAs and change their connectivity unpredictably.

\begin{figure}
\begin{centering}
\includegraphics[width=0.5\columnwidth]{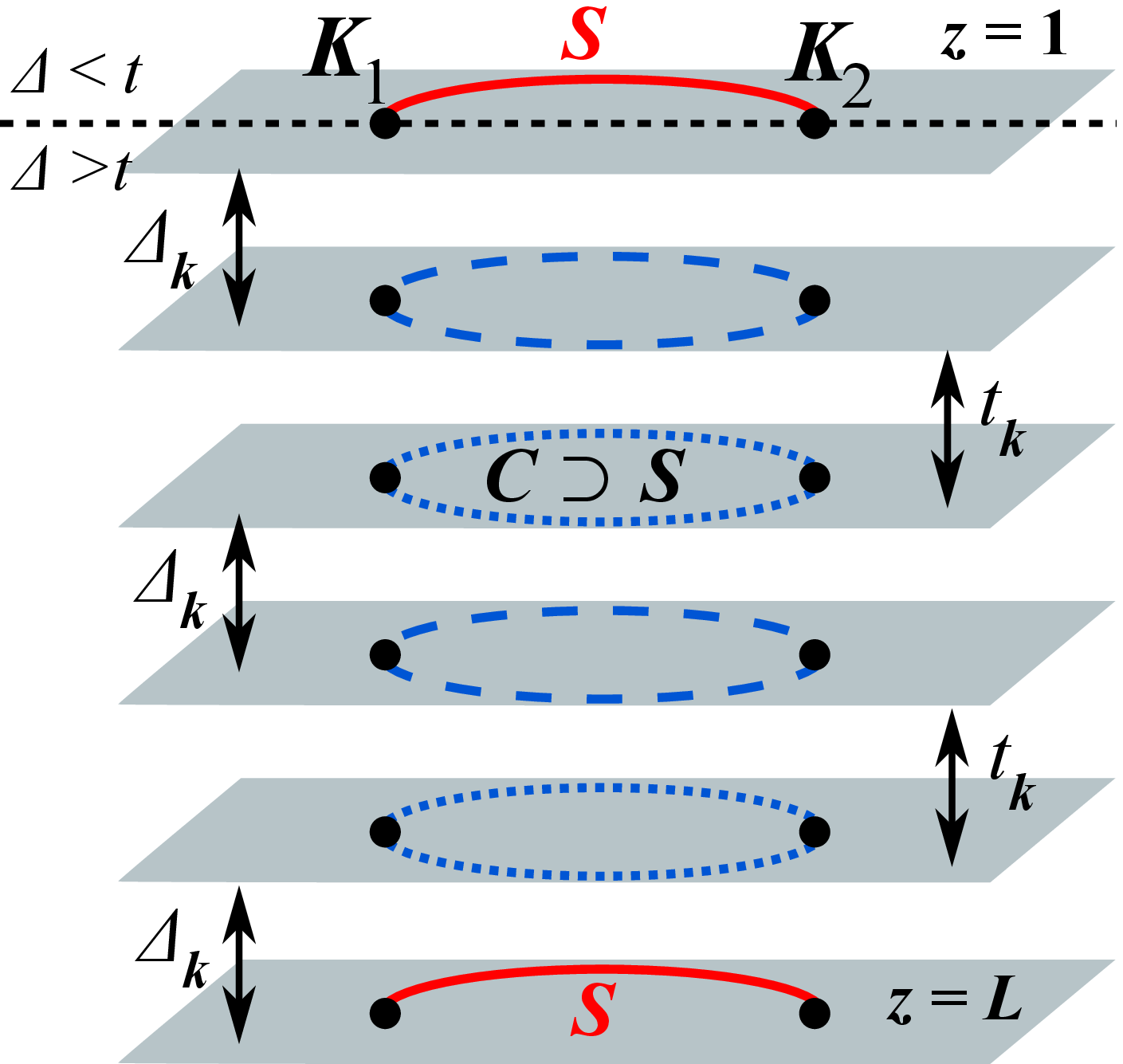}\includegraphics[width=0.5\columnwidth]{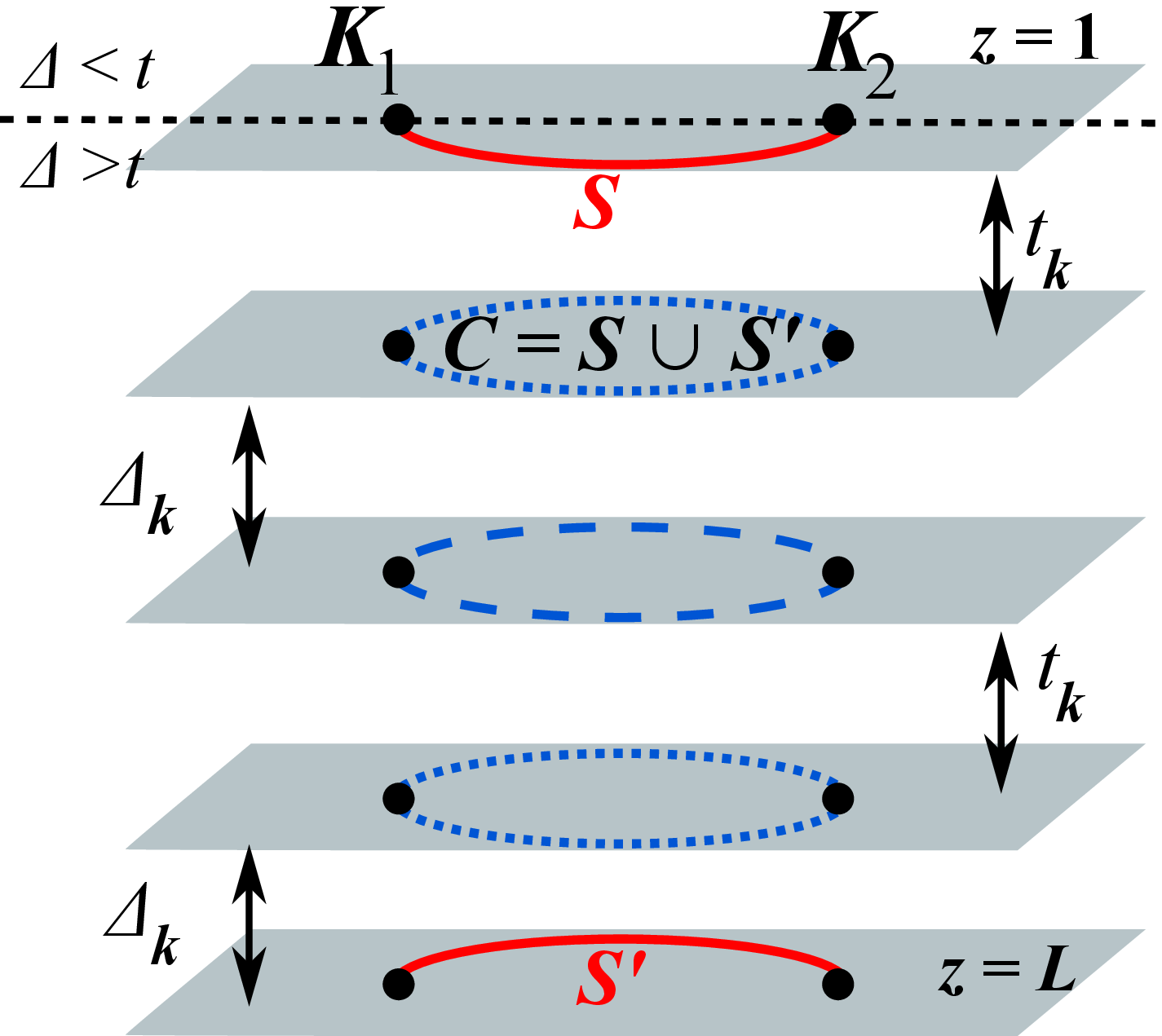}
\par\end{centering}

\rule[0.5ex]{1\columnwidth}{1pt}

\begin{centering}
\includegraphics[height=1.5cm]{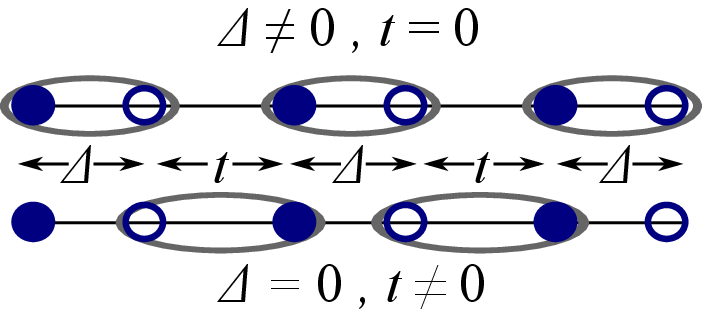}~~~~~~~\includegraphics[height=1.5cm]{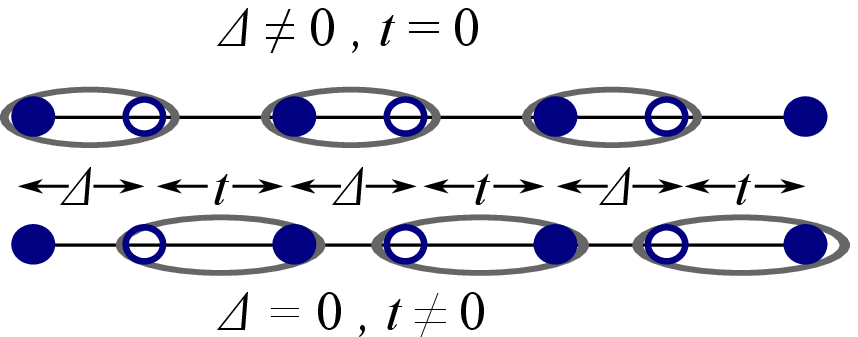}
\par\end{centering}

\caption{Above: Layering prescription for obtaining FAs of arbitrary shape.
Dotted (dashed) ellipses represent electron (hole) FSs, and solid
red segments are the residual FAs on adding interlayer hoppings $t_{\boldsymbol{k}}$
and $\Delta_{\boldsymbol{k}}$. The horizontal black dashed line on
the topmost layer separates regions with $\Delta_{\boldsymbol{k}}>t_{\boldsymbol{k}}$
and $\Delta_{\boldsymbol{k}}<t_{\boldsymbol{k}}$. An even (odd) number
of total layers gives identical (non-identical) FAs on the two surfaces,
as shown on the left (right). Below: 1D systems at fixed $\boldsymbol{k}\in C$
under the influence of $\Delta$ and $t$ in the extreme cases where
the smaller hopping vanishes, for even (left) and odd (right) $L$.
Filled (empty) circles denote a state on an electron (a hole) Fermi
surface in the limit of decoupled layers. The ellipses enclose the
states which get mutually gapped out by the hoppings.\label{fig:layering picture}}
\end{figure}

\textbf{Layering prescription:} We describe the prescription by considering
the simplest WSM, which has just two Weyl nodes, at $(k_{x},k_{y},k_{z})\equiv(\boldsymbol{k},k_{z})=(\boldsymbol{K}_{1,2},0)$
and the FAs connect $\boldsymbol{K}_{1}$ and $\boldsymbol{K}_{2}$
along a segment $S$ ($S'$) on the $z=1$ ($z=L$) surface in the
surface Brillouin zone, as shown in Fig \ref{fig:layering picture}.
$S=S'$ and $S\neq S'$ correspond to the two qualitatively different
situations (a) and (b) mentioned earlier, and will be obtained by
distinct boundary conditions. Generalization to more Weyl points and
Weyl points away from the $k_{z}=0$ plane is straightforward.

We claim that this WSM is generated by the following Bloch Hamiltonian:
\begin{eqnarray}
H_{\boldsymbol{k}} & = & \sum_{z=1}^{L}\psi_{z,\boldsymbol{k}}^{\dagger}(-1)^{z}\mathcal{E}_{\boldsymbol{k}}\psi_{z,\boldsymbol{k}}\label{eq:main H}\\
 &  & +\sum_{z=1}^{L-1}\psi_{z,\boldsymbol{k}}^{\dagger}h_{z,\boldsymbol{k}}\psi_{z+1,\boldsymbol{k}}+\mbox{h.c.}\nonumber 
\end{eqnarray}
where even (odd) $L$ generates $S=S'$ ($S\neq S'$), $\mathcal{E}_{\boldsymbol{k}}$
is a phenomenological function that vanishes along a contour $C$
given by
\begin{equation}
C\begin{cases}
\supset S & S=S'\\
=S\cup S' & S\neq S'
\end{cases}\label{eq:contour}
\end{equation}
and the interlayer coupling $h_{z,\boldsymbol{k}}=-t_{\boldsymbol{k}}(\Delta_{\boldsymbol{k}})$
if $z$ is even (odd). If $S=S'$, $C$ can be chosen arbitrarily
as long as it contains the entire segment $S$. The functions $t_{\boldsymbol{k}}$
and $\Delta_{\boldsymbol{k}}$ are real, non-negative phenomenological
functions that satisfy:
\begin{equation}
t_{\boldsymbol{k}}\begin{cases}
>\Delta_{\boldsymbol{k}} & \boldsymbol{k}\in S\\
<\Delta_{\boldsymbol{k}} & \boldsymbol{k}\in C/S(=S'\mbox{ if }C=S\cup S')
\end{cases}\label{eq:tDelta condition}
\end{equation}
(\ref{eq:tDelta condition}) dictates that $t_{\boldsymbol{k}}=\Delta_{\boldsymbol{k}}$
exactly at $\boldsymbol{k}=\boldsymbol{K}_{1,2}$. The $\boldsymbol{k}$-dependence
of $t$ and $\Delta$ away from $C$ is unimportant for our purposes,
and will be assumed to be negligible henceforth. We now justify the
above claim.

\emph{Bulk:}\textbf{\emph{ }}If the interlayer couplings $t_{\boldsymbol{k}}=\Delta_{\boldsymbol{k}}=0$,
then $\mathbb{H}=\sum_{\boldsymbol{k}}H_{\boldsymbol{k}}$ describes
a stack of alternate \emph{non-degenerate} electron and hole FSs.
When $t_{\boldsymbol{k}}$ and $\Delta_{\boldsymbol{k}}$ are turned
on, these FSs get gapped out in pairs in the bulk. Indeed, the bulk
Hamiltonian is 
\begin{equation}
H_{\boldsymbol{k},k_{z}}^{\mbox{bulk}}=\mathcal{E}_{\boldsymbol{k}}\sigma_{z}+\left(\Delta_{\boldsymbol{k}}-t_{\boldsymbol{k}}\cos k_{z}\right)\sigma_{x}+t_{\boldsymbol{k}}\sin k_{z}\sigma_{y}
\end{equation}
where $z$ is the layering direction, and is gapped everywhere except
at the desired Weyl points: $(\boldsymbol{k},k_{z})=\left(\boldsymbol{K}_{1,2},0\right)$,
due to (\ref{eq:tDelta condition}). Allowing $t_{\boldsymbol{k}}$
and $\Delta_{\boldsymbol{k}}$ to be negative or complex simply moves
the Weyl points off the $k_{z}=0$ plane, but this does not affect
the shape of the FAs. Near the gapless points, $H^{\mbox{bulk}}$
realizes the Weyl Hamiltonian: 
\begin{align}
 & H_{\boldsymbol{K}_{i}+\boldsymbol{p},0+p_{z}}^{\mbox{bulk}}\nonumber \\
 & \,\,\,\approx\left[\boldsymbol{p}\cdotp\boldsymbol{\nabla_{k}}\mathcal{E}_{\boldsymbol{K}_{i}}\right]\sigma_{z}+\left[\boldsymbol{p}\cdot\boldsymbol{\nabla_{k}}\left(\Delta_{\boldsymbol{K}_{i}}-t_{\boldsymbol{K}_{i}}\right)\right]\sigma_{x}+\left[t_{\boldsymbol{K}_{i}}p_{z}\right]\sigma_{y}\nonumber \\
 & \,\,\,\equiv p_{\perp}v_{F}(\boldsymbol{K}_{i})\sigma_{z}+p_{\parallel}v_{i}\sigma_{x}+\Delta_{0}p_{z}\sigma_{y}
\end{align}
where $p_{\perp}=\boldsymbol{p}\cdot\hat{\boldsymbol{e}}_{r}(\boldsymbol{K}_{i})$
and $p_{\parallel}=\boldsymbol{p}\cdot\hat{\boldsymbol{e}}_{t}(\boldsymbol{K}_{i})$
are momenta perpendicular and parallel to $C$ ($\hat{e}_{r}(\boldsymbol{k})$
and $\hat{\boldsymbol{e}}_{t}(\boldsymbol{k})$ are 2D unit vectors
normal and tangential to $C$), $v_{F}$ is the Fermi velocity of
the 2D FSs and $v_{i}=\boldsymbol{\nabla_{k}}\left(\Delta_{\boldsymbol{K}_{i}}-t_{\boldsymbol{K}_{i}}\right)\cdot\hat{e_{t}}(\boldsymbol{K}_{i})$.
In going to the second line, the variation of $t_{\boldsymbol{k}}$
and $\Delta_{\boldsymbol{k}}$ perpendicular to $C$ has been assumed
to be negligible, since it does not affect the shape of the FAs. $v_{i}$
has opposite signs at $\boldsymbol{K}_{1}$ and $\boldsymbol{K}_{2}$,
ensuring that the Weyl nodes have opposite chirality. $H_{\mbox{bulk}}$
is obviously unaffected by the boundary conditions at $z=1$ and $z=L$
for large $L$.

\emph{Surface:} The surface, however, strongly depends on the boundary
conditions; in particular, it is qualitatively different for odd and
even $L$. If $L$ is odd, at each $\boldsymbol{k}\in C$, a state
remains unpaired and hence, gapless, at $z=1$ ($z=L$) whenever $\Delta_{\boldsymbol{k}}<t_{\boldsymbol{k}}$
($\Delta_{\boldsymbol{k}}>t_{\boldsymbol{k}}$). The gapless states
at $z=1$ ($z=L$) thus, trace out $S$ ($S'$). On the other hand,
when $L$ is even, both ends of a the chain at fixed $\boldsymbol{k}$
carry a gapless state when $\Delta_{\boldsymbol{k}}<t_{\boldsymbol{k}}$
and neither end has gapless states when $\Delta_{\boldsymbol{k}}>t_{\boldsymbol{k}}$.
In this case, the gapless states on both surfaces of the slab trace
out $S$. 

Viewed differently, the 1D system at fixed $\boldsymbol{k}\in C$
and $\left|\Delta\right|\neq\left|t\right|$ is an insulator in the
CII symmetry class, which is known to have a $\mathbb{Z}$ topological
classification \cite{SFRLClassification,KitaevClassification}. While
$\left|\Delta\right|>\left|t\right|$ gives a trivial phase, $\left|\Delta\right|<\left|t\right|$
is topologically non-trivial with a zero mode at each end protected
by a chiral symmetry, if the 1D lattice has a whole number of unit
cells. These end states are nothing but the FA states at that $\boldsymbol{k}$
when $L$ is even. As $\boldsymbol{k}$ is varied along $C$, the
1D system undergoes a topological phase transition at $\boldsymbol{K}_{1}$
and $\boldsymbol{K}_{2}$. For odd $L$, there is always a state at
one end of the chain, as show in Fig \ref{fig:layering picture}.
This prescription is similar in spirit to node-pairing picture of
Ref \cite{HosurRyuChiralTISC} for chiral topological insulators in
three dimensions. Note that it is necessary to start with two sets
of FSs, since a single FS cannot be destroyed perturbatively.

\emph{Symmetry analysis: }WSMs can only exist in systems in which
at least one symmetry out of time-reversal symmetry ($\mathcal{T}$)
and inversion symmetry ($\mathcal{I}$) is broken; the presence of
both would make each band doubly degenerate and give Dirac semi-metals
instead with four-component fermions near at each node. Moreover,
$\mathcal{I}$-symmetric, $\mathcal{T}$-breaking ($\mathcal{T}$-symmetric,
$\mathcal{I}$-breaking) WSMs have an odd (even) number of pairs of
Weyl nodes. In the current picture, the breaking of these symmetries
can be understood as follows. Let us assume $\mathcal{E}_{\boldsymbol{k}}=\mathcal{E}_{-\boldsymbol{k}}$;
if this weren't true, both symmetries would broken from the outset.
In general, $t_{\boldsymbol{k}}$ and $\Delta_{\boldsymbol{k}}$ are
unrelated to $t_{-\boldsymbol{k}}$ and $\Delta_{-\boldsymbol{k}}$,
in which case both symmetries would again be broken. However, $\mathcal{T}$
is preserved if $t_{\boldsymbol{k}}=t_{-\boldsymbol{k}}$ and $\Delta_{\boldsymbol{k}}=\Delta_{-\boldsymbol{k}}$,
which can only happen if the number of points on $C$ at which $t_{\boldsymbol{k}}=\Delta_{\boldsymbol{k}}$,
and thus the number of Weyl nodes, is an integer multiple of four.
On the other hand, inversion about a particular layer interchanges
$t$ and $\Delta$; thus, $t_{\boldsymbol{k}}=\Delta_{-\boldsymbol{k}}$
preserves this inversion symmetry. This condition requires $t_{\boldsymbol{k}}-\Delta_{\boldsymbol{k}}$
to change sign twice an odd number of times along $C$, giving an
odd number of Weyl node pairs.

\begin{figure}
~

~

~

\includegraphics[width=0.16\columnwidth]{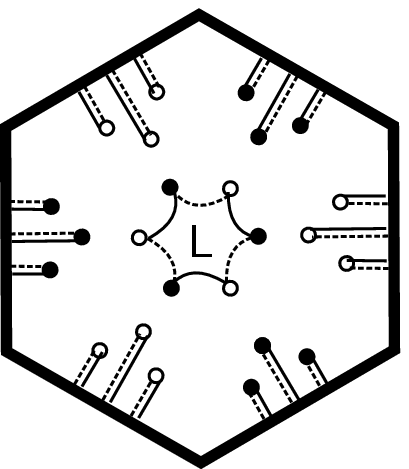}\includegraphics[width=0.42\columnwidth]{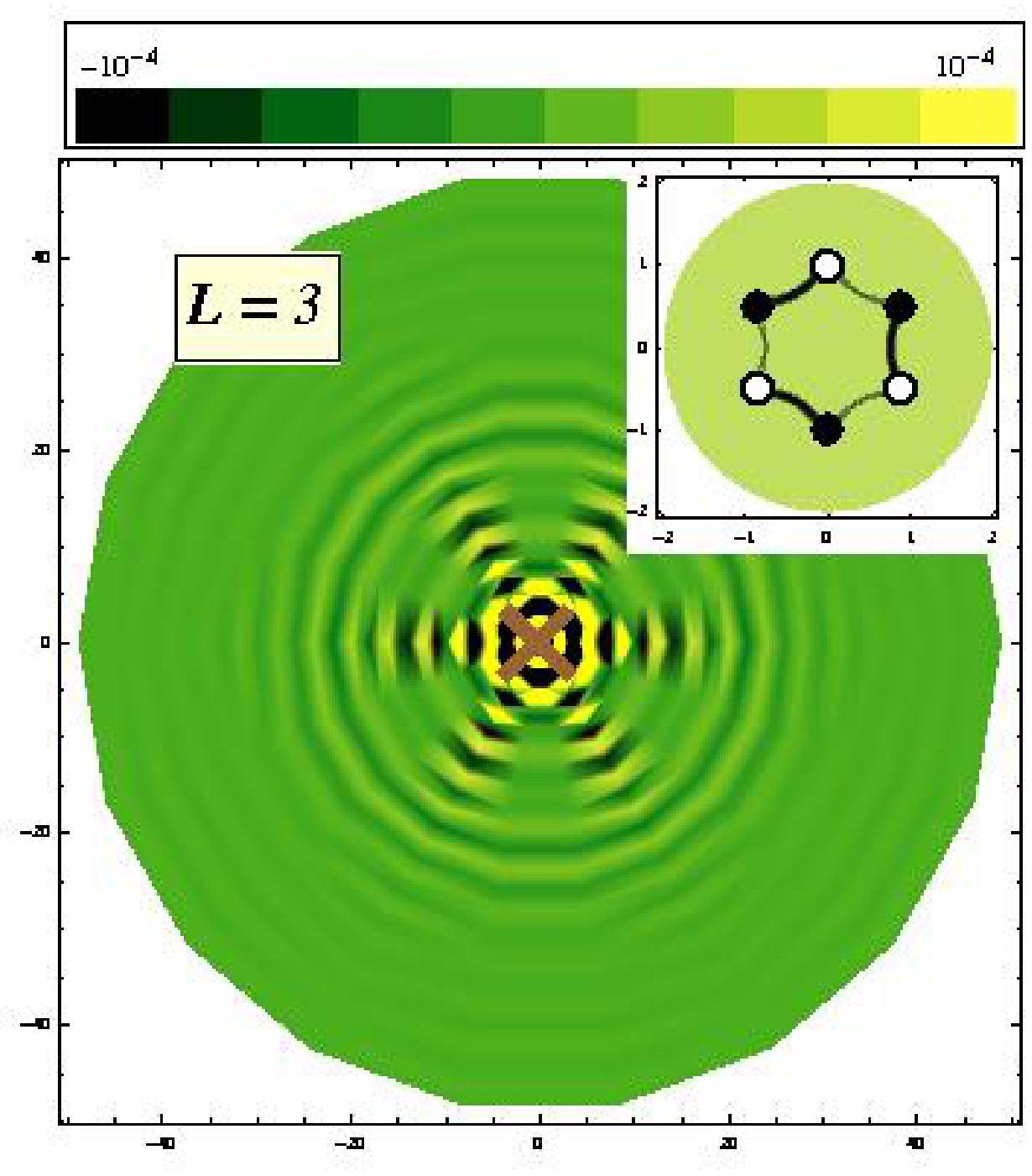}\includegraphics[width=0.42\columnwidth]{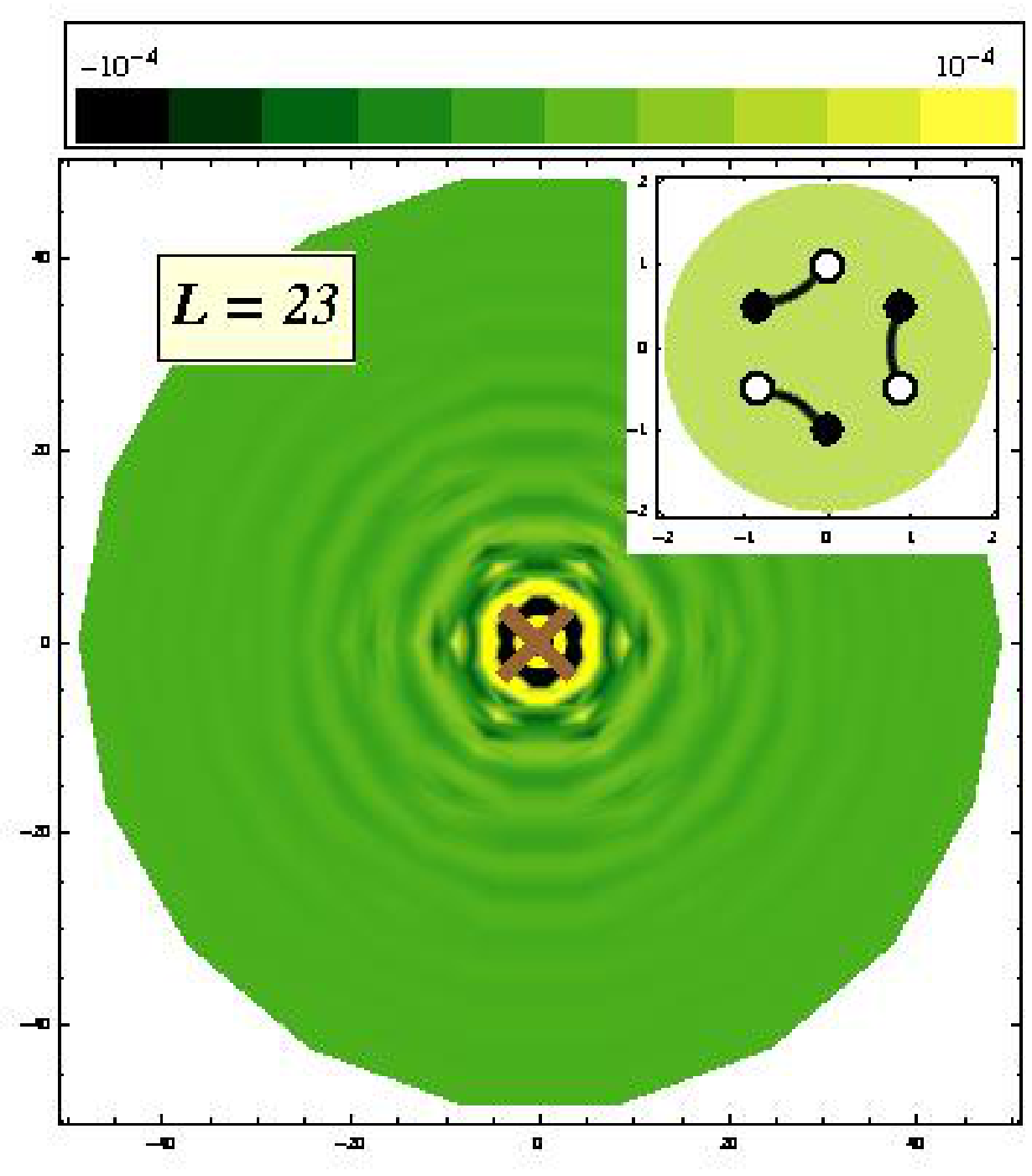}

\begin{raggedright}
~~~~~~~(a)~~~~~~~~~~~~~~~~~~~~~~~~~~~~~~(b)~~~~~~~~~~~~~~~~~~~~~~~~~~~~~~~~~~~~~~~~~~~~(c)~~~~~~~~~~~~~~~~~~~~~~~~~~~~~~
\par\end{raggedright}

\caption{(a) FAs in the surface Brillouin zone on the 111 surface of iridate
WSMs. Solid (dashed) lines denote FAs on the near (far) surface. Filled
(empty) circles denote projections of Weyl nodes of positive (negative)
chirality in all the figures above. (b and c) Surface LDOS in arbitrary
units due to the six FAs near the $L$ point in the presence of a
point scatterer on the surface (brown cross) for a thin sample (b)
and a thick sample (c). Insets show the numerically computed surface
spectral function at $E=0$ for the clean system, with darker colors
representing larger values. The computation is done for the model,
described in the text, which generates the six FAs near the $L$-point
but not the remaining eighteen FAs near the Brillouin zone edges.\label{fig:iridate STM}}
\end{figure}

\textbf{LDOS results:} Having described the procedure for obtaining
FAs from a 2D limit, we demonstrate its utility by calculating the
surface spectral function for a clean system and the surface LDOS
in the presence of a point surface scatterer within a model that should
be relevant to the pyrochlore iridates A$_{2}$Ir$_{2}$O$_{7}$,
A$=$Y, Eu, which are purported WSMs with 24 Weyl nodes \cite{PyrochloreWeyl}.
The lattice in the WSM phase has inversion symmetry as well as a threefold
rotation symmetry $R_{3}$ about the cubic 111-axis, and there are
six Weyl nodes related by these symmetries near each of the four $L$
points in the FCC Brillouin zone. Additionally, the lattice also has
a $D_{6}$ symmetry, i.e., $\pi/3$ rotation about {[}111{]} followed
by reflection in the perpendicular plane. This symmetry has an important
implication for the FAs: if it is preserved in a slab geometry, then
the FAs will be as shown in Fig \ref{fig:iridate STM} (a). In particular,
the six FAs near the center of the surface Brillouin zone enclose
an area, while the remaining eighteen overlap in pairs on the opposite
surfaces. We note that Ref \cite{KrempaWeyl} also predicted Weyl
semimetallic behavior in the above iridates, but with only 8 Weyl
nodes. In this case, the hexagonal figure around the $L$ point would
collapse to a point. As we argue below, LDOS oscillations on the surface
stem predominantly from the hexagonal figure; hence, the proposal
of Ref \cite{KrempaWeyl}, if true, would imply no strong LDOS oscillations.

We compute the surface LDOS for a model that has six FAs like the
ones around the $L$ point, as a function of the the sample thickness.
The remaining eighteen FAs in the iridates are expected to be destroyed
by finite size effects for thin samples, while for thick samples backscattering
occurs across the Brillouin zone and hence can give rise to LDOS oscillations
only on the lattice scale. These oscillations are unlikely to be distinguishable
from the electron density variations on this scale already present.

The LDOS is calculated via the standard $T$-matrix formalism. Given
the time-ordered Green's function for the clean system: $G_{0}(\boldsymbol{k},E)=\left(E-H_{\boldsymbol{k}}\right)^{-1}$
and a scattering potential: $U_{z,z'}(x,y)=u\delta(x)\delta(y)\delta_{z,1}\delta_{z',1}$,
the $T$-matrix is given by $T(\omega)=\left(1-U\sum_{\boldsymbol{k}}G_{0}(\boldsymbol{k},E)\right)^{-1}U$,
independent of momentum, since the scattering potential in momentum
independent. Here, $G_{0}$, $U$ and $T$ are all $L\times L$ matrices
indexed by $z$. The full Green's function in the presence of the
impurity is $G(\boldsymbol{k},\boldsymbol{k}',E)=\delta_{\boldsymbol{k},\boldsymbol{k}'}G_{0}(\boldsymbol{k},E)+G_{0}(\boldsymbol{k},E)T(\omega)G_{0}(\boldsymbol{k}',E)$,
and the LDOS on the $z=1$ surface is related to the $(1,1)$ element
of its retarded cousin: $\rho_{\mbox{surf}}(\boldsymbol{r},E)=-\frac{1}{\pi}\mathrm{Im}G^{11}(\boldsymbol{r},\boldsymbol{r},E+i\delta)$,
where $G(\boldsymbol{r},\boldsymbol{r}',E)=\int_{\boldsymbol{k},\boldsymbol{k}'}e^{i(\boldsymbol{k}\cdot\boldsymbol{r}-\boldsymbol{k}'\cdot\boldsymbol{r}')}G(\boldsymbol{k},\boldsymbol{k}',E)$.
For the calculation, we use $\mathcal{E}(\boldsymbol{k})=\sqrt{k^{2}+2k^{6}\cos^{2}3\theta_{\boldsymbol{k}}}-1$
to generate the hexagonal figure and $t_{\boldsymbol{k}}\equiv t_{0}=0.5$,
$\Delta_{\boldsymbol{k}}=t_{0}(1-\cos3\theta_{\boldsymbol{k}})$ to
obtain the FAs. Here, $(k,\theta_{\boldsymbol{k}})$ the polar coordinates
of $\boldsymbol{k}$.

The results are presented in Fig \ref{fig:iridate STM} (b) and (c)
for $E=0$. For thin samples, clear LDOS oscillations are seen in
the horizontal direction as well as along the two equivalent directions
related by $\pi/3$ rotation. The origin of these oscillations becomes
clear if one looks at $A_{\mbox{surf}}^{0}(\boldsymbol{k},E=0)$,
displayed inset. Since the sample is thin, FA wavefunctions from the
far surface have significant amplitude on the near surface, which
allows backscattering to occur diametrically across the hexagon. The
dominant backscattering processes are the ones involving the midpoints
of the FAs, since the Fermi surface is nested here. On the other hand,
as the sample thickness is increased, three of the six FAs retreat
to the far surface and backscattering is exponentially suppressed.
The result is small variations in the LDOS arising from scattering
between FAs solely on the top surface. Thus, the STM map has a distinct
evolution with sample thickness which is characteristic of the FA
structure in the iridate WSMs.

\textbf{Surface spectral function in clean thermodynamic limit:} $A_{\mbox{surf}}^{0}(\boldsymbol{k},E)$
was computed numerically in order to generate the insets of Fig \ref{fig:iridate STM}.
In the thermodynamic limit, however, this calculation can be done
analytically. Denoting the $(1,1)$ element of the clean Green's function
for a $L$-layer slab by $G_{0(L)}^{11}$, it is straightforward to
show, by explicitly evaluating the $(1,1)$ cofactor of $E-H_{\boldsymbol{k}}$
and using $\mathrm{det}\left(\begin{array}{cc}
A & B\\
C & D
\end{array}\right)=\mathrm{det}(A)\mathrm{det}(D-CA^{-1}B)$, that
\begin{align}
 & (E-\mathcal{E}_{\boldsymbol{k}})G_{0(L)}^{11}(\boldsymbol{k},E)\nonumber \\
 & =1+\frac{\Delta_{\boldsymbol{k}}^{2}}{E^{2}-\mathcal{E}_{\boldsymbol{k}}^{2}-\Delta_{\boldsymbol{k}}^{2}-t_{\boldsymbol{k}}^{2}(E-\mathcal{E}_{\boldsymbol{k}})G_{0(L-2)}^{11}(\boldsymbol{k},E)}
\end{align}
In the thermodynamic limit: $L\to\infty$, $G_{0(L)}^{11}(\boldsymbol{k},E)\approx G_{0(L-2)}^{11}(\boldsymbol{k},E)\equiv g(\boldsymbol{k},E)$.
Thus,\begin{widetext} 
\begin{equation}
g(\boldsymbol{k},E)=\frac{1}{2t_{\boldsymbol{k}}^{2}(E-\mathcal{E}_{\boldsymbol{k}})}\left[\left(E^{2}-\mathcal{E}_{\boldsymbol{k}}^{2}+t_{\boldsymbol{k}}^{2}-\Delta_{\boldsymbol{k}}^{2}\right)\pm\sqrt{\left(E^{2}-\mathcal{E}_{\boldsymbol{k}}^{2}+t_{\boldsymbol{k}}^{2}-\Delta_{\boldsymbol{k}}^{2}\right)^{2}-4t_{\boldsymbol{k}}^{2}(E^{2}-\mathcal{E}_{\boldsymbol{k}}^{2})}\right]\label{eq:gtherm}
\end{equation}
\end{widetext}The physical condition $A_{\mbox{surf}}^{0}(\boldsymbol{k},E)=-\frac{1}{\pi}\mathrm{Im}\left[g(\boldsymbol{k},E+i\delta)\right]\ge0$
fixes the sign in front of the square root. Clearly, 
\begin{gather}
A_{\mbox{surf}}^{0}(\boldsymbol{k},E)=\delta(E-\mathcal{E}_{\boldsymbol{k}})\frac{t_{\boldsymbol{k}}^{2}-\Delta_{\boldsymbol{k}}^{2}+\left|t_{\boldsymbol{k}}^{2}-\Delta_{\boldsymbol{k}}^{2}\right|}{2t_{\boldsymbol{k}}^{2}}+\nonumber \\
\frac{1}{2t_{\boldsymbol{k}}^{2}(E-\mathcal{E}_{\boldsymbol{k}})}\mathrm{Im}\sqrt{\left(E^{2}-\mathcal{E}_{\boldsymbol{k}}^{2}+t_{\boldsymbol{k}}^{2}-\Delta_{\boldsymbol{k}}^{2}\right)^{2}-4t_{\boldsymbol{k}}^{2}(E^{2}-\mathcal{E}_{\boldsymbol{k}}^{2})}\label{eq:asurf}
\end{gather}
The first line is non-zero only when $t_{\boldsymbol{k}}^{2}>\Delta_{\boldsymbol{k}}^{2}$
and has a sharp peak at $E=\mathcal{E}_{\boldsymbol{k}}$. Clearly,
this represents the contribution to $A_{\mbox{surf}}^{0}$ from the
FA. Whereas, the second line is non-vanishing when $|E|>|\mathcal{E}_{\boldsymbol{k}}|$
and $\left|t_{\boldsymbol{k}}-\sqrt{E^{2}-\mathcal{E}_{\boldsymbol{k}}^{2}}\right|<|\Delta_{\boldsymbol{k}}|<\left|t_{\boldsymbol{k}}+\sqrt{E^{2}-\mathcal{E}_{\boldsymbol{k}}^{2}}\right|$.
These inequalities are satisfied in the region near the projection
of the Weyl points onto the surface. Moreover, this contribution to
$A_{\mbox{surf}}^{0}$ has no delta-function peak. Thus, it represents
contributions form the bulk states near the Weyl nodes. The quantity
$A_{\mbox{surf}}^{0}$ can be directly measured by ARPES experiments.

In conclusion, we have studied impurity-induced Friedel oscillations
due to FAs in WSMs, focusing on the FA structure of the purported
iridate WSMs, and observed their dependence on sample thickness. For
thin samples, FAs on both surfaces collude to allow nested backscattering
and hence produce strong LDOS oscillations, whereas for thick samples,
the FAs on the far surface do not reach the near surface and such
backscattering and the consequent LDOS oscillations are suppressed.
The calculation is done by building the desired WSM and FA structure
by stacking electron and hole Fermi surfaces and adding suitable interlayer
hopping. Within this prescription, the surface spectral function for
a clean system can be calculated analytically in the thermodynamic
limit.
\begin{acknowledgments}
P.H. would like to thank Ashvin Vishwanath, Tarun Grover, Siddharth
Parameswaran, Xiaoliang Qi and Shou-Cheng Zhang for insightful discussions,
Ashvin Vishwanath and Tarun Grover for useful comments on the manuscript,
and DOE grant DE-AC02-05CH11231 for financial support during most
of this work. During the final stages of writing this manuscript,
P.H. was supported by Packard Fellowship grant 1149927-100-UAQII.
\end{acknowledgments}
\bibliographystyle{apsrev}
\bibliography{references}

\begin{thebibliography}{27}
\expandafter\ifx\csname natexlab\endcsname\relax\def\natexlab#1{#1}\fi
\expandafter\ifx\csname bibnamefont\endcsname\relax
  \def\bibnamefont#1{#1}\fi
\expandafter\ifx\csname bibfnamefont\endcsname\relax
  \def\bibfnamefont#1{#1}\fi
\expandafter\ifx\csname citenamefont\endcsname\relax
  \def\citenamefont#1{#1}\fi
\expandafter\ifx\csname url\endcsname\relax
  \def\url#1{\texttt{#1}}\fi
\expandafter\ifx\csname urlprefix\endcsname\relax\def\urlprefix{URL }\fi
\providecommand{\bibinfo}[2]{#2}
\providecommand{\eprint}[2][]{\url{#2}}

\bibitem[{\citenamefont{Volovik}(2009)}]{VolovikBook}
\bibinfo{author}{\bibfnamefont{G.}~\bibnamefont{Volovik}},
  \emph{\bibinfo{title}{The Universe in a Helium Droplet}}, International
  Series of Monographs on Physics (\bibinfo{publisher}{Oxford University
  Press}, \bibinfo{year}{2009}), ISBN \bibinfo{isbn}{9780199564842}.

\bibitem[{\citenamefont{Wan et~al.}(2011)\citenamefont{Wan, Turner, Vishwanath,
  and Savrasov}}]{PyrochloreWeyl}
\bibinfo{author}{\bibfnamefont{X.}~\bibnamefont{Wan}},
  \bibinfo{author}{\bibfnamefont{A.~M.} \bibnamefont{Turner}},
  \bibinfo{author}{\bibfnamefont{A.}~\bibnamefont{Vishwanath}},
  \bibnamefont{and} \bibinfo{author}{\bibfnamefont{S.~Y.}
  \bibnamefont{Savrasov}}, \bibinfo{journal}{Phys. Rev. B}
  \textbf{\bibinfo{volume}{83}}, \bibinfo{pages}{205101}
  (\bibinfo{year}{2011}).

\bibitem[{\citenamefont{Witczak-Krempa and Kim}(2012)}]{KrempaWeyl}
\bibinfo{author}{\bibfnamefont{W.}~\bibnamefont{Witczak-Krempa}}
  \bibnamefont{and} \bibinfo{author}{\bibfnamefont{Y.-B.} \bibnamefont{Kim}},
  \bibinfo{journal}{Phys. Rev. B} \textbf{\bibinfo{volume}{85}},
  \bibinfo{pages}{045124} (\bibinfo{year}{2012}).

\bibitem[{\citenamefont{Nielsen and
  Ninomiya}(1981{\natexlab{a}})}]{NielsenFermionDoubling1}
\bibinfo{author}{\bibfnamefont{H.}~\bibnamefont{Nielsen}} \bibnamefont{and}
  \bibinfo{author}{\bibfnamefont{M.}~\bibnamefont{Ninomiya}},
  \bibinfo{journal}{Nuclear Physics B} \textbf{\bibinfo{volume}{185}},
  \bibinfo{pages}{20 } (\bibinfo{year}{1981}{\natexlab{a}}), ISSN
  \bibinfo{issn}{0550-3213}.

\bibitem[{\citenamefont{Nielsen and
  Ninomiya}(1981{\natexlab{b}})}]{NielsenFermionDoubling2}
\bibinfo{author}{\bibfnamefont{H.~B.} \bibnamefont{Nielsen}} \bibnamefont{and}
  \bibinfo{author}{\bibfnamefont{M.}~\bibnamefont{Ninomiya}},
  \bibinfo{journal}{Nuclear Physics B} \textbf{\bibinfo{volume}{193}},
  \bibinfo{pages}{173} (\bibinfo{year}{1981}{\natexlab{b}}).

\bibitem[{\citenamefont{Burkov and Balents}(2011)}]{WeylMultiLayer}
\bibinfo{author}{\bibfnamefont{A.~A.} \bibnamefont{Burkov}} \bibnamefont{and}
  \bibinfo{author}{\bibfnamefont{L.}~\bibnamefont{Balents}},
  \bibinfo{journal}{Phys. Rev. Lett.} \textbf{\bibinfo{volume}{107}},
  \bibinfo{pages}{127205} (\bibinfo{year}{2011}).

\bibitem[{\citenamefont{{Fang} et~al.}(2011)\citenamefont{{Fang}, {Gilbert},
  {Dai}, and {Bernevig}}}]{BernevigDoubleWeyl}
\bibinfo{author}{\bibfnamefont{C.}~\bibnamefont{{Fang}}},
  \bibinfo{author}{\bibfnamefont{M.~J.} \bibnamefont{{Gilbert}}},
  \bibinfo{author}{\bibfnamefont{X.}~\bibnamefont{{Dai}}}, \bibnamefont{and}
  \bibinfo{author}{\bibfnamefont{B.~A.} \bibnamefont{{Bernevig}}},
  \bibinfo{journal}{ArXiv e-prints}  (\bibinfo{year}{2011}),
  \eprint{1111.7309}.

\bibitem[{\citenamefont{Delplace et~al.}(2012)\citenamefont{Delplace, Li, and
  Carpentier}}]{CarpentierWeyl}
\bibinfo{author}{\bibfnamefont{P.}~\bibnamefont{Delplace}},
  \bibinfo{author}{\bibfnamefont{J.}~\bibnamefont{Li}}, \bibnamefont{and}
  \bibinfo{author}{\bibfnamefont{D.}~\bibnamefont{Carpentier}},
  \bibinfo{journal}{EPL (Europhysics Letters)} \textbf{\bibinfo{volume}{97}},
  \bibinfo{pages}{67004} (\bibinfo{year}{2012}).

\bibitem[{\citenamefont{{Cho}}(2011)}]{ChoTItoWeyl}
\bibinfo{author}{\bibfnamefont{G.~Y.} \bibnamefont{{Cho}}},
  \bibinfo{journal}{ArXiv e-prints}  (\bibinfo{year}{2011}),
  \eprint{1110.1939}.

\bibitem[{\citenamefont{Hal\'asz and Balents}(2012)}]{HalaszWeyl}
\bibinfo{author}{\bibfnamefont{G.~B.} \bibnamefont{Hal\'asz}} \bibnamefont{and}
  \bibinfo{author}{\bibfnamefont{L.}~\bibnamefont{Balents}},
  \bibinfo{journal}{Phys. Rev. B} \textbf{\bibinfo{volume}{85}},
  \bibinfo{pages}{035103} (\bibinfo{year}{2012}).

\bibitem[{\citenamefont{Jiang}(2012)}]{JiangWeyl}
\bibinfo{author}{\bibfnamefont{J.-H.} \bibnamefont{Jiang}},
  \bibinfo{journal}{Phys. Rev. A} \textbf{\bibinfo{volume}{85}},
  \bibinfo{pages}{033640} (\bibinfo{year}{2012}).

\bibitem[{\citenamefont{{Lu} et~al.}(2012)\citenamefont{{Lu}, {Fu},
  {Joannopoulos}, and {Solja{\v c}i{\'c}}}}]{PhotonicCrystalWeyl}
\bibinfo{author}{\bibfnamefont{L.}~\bibnamefont{{Lu}}},
  \bibinfo{author}{\bibfnamefont{L.}~\bibnamefont{{Fu}}},
  \bibinfo{author}{\bibfnamefont{J.~D.} \bibnamefont{{Joannopoulos}}},
  \bibnamefont{and} \bibinfo{author}{\bibfnamefont{M.}~\bibnamefont{{Solja{\v
  c}i{\'c}}}}, \bibinfo{journal}{ArXiv e-prints}  (\bibinfo{year}{2012}),
  \eprint{1207.0478}.

\bibitem[{\citenamefont{Yang et~al.}(2011)\citenamefont{Yang, Lu, and
  Ran}}]{RanQHWeyl}
\bibinfo{author}{\bibfnamefont{K.-Y.} \bibnamefont{Yang}},
  \bibinfo{author}{\bibfnamefont{Y.-M.} \bibnamefont{Lu}}, \bibnamefont{and}
  \bibinfo{author}{\bibfnamefont{Y.}~\bibnamefont{Ran}},
  \bibinfo{journal}{Phys. Rev. B} \textbf{\bibinfo{volume}{84}},
  \bibinfo{pages}{075129} (\bibinfo{year}{2011}).

\bibitem[{\citenamefont{Xu et~al.}(2011)\citenamefont{Xu, Weng, Wang, Dai, and
  Fang}}]{FangChernSemimetal}
\bibinfo{author}{\bibfnamefont{G.}~\bibnamefont{Xu}},
  \bibinfo{author}{\bibfnamefont{H.~M.} \bibnamefont{Weng}},
  \bibinfo{author}{\bibfnamefont{Z.}~\bibnamefont{Wang}},
  \bibinfo{author}{\bibfnamefont{X.}~\bibnamefont{Dai}}, \bibnamefont{and}
  \bibinfo{author}{\bibfnamefont{Z.}~\bibnamefont{Fang}},
  \bibinfo{journal}{Phys. Rev. Lett.} \textbf{\bibinfo{volume}{107}},
  \bibinfo{pages}{186806} (\bibinfo{year}{2011}).

\bibitem[{\citenamefont{Nielsen and Ninomiya}(1983)}]{NielsenABJ}
\bibinfo{author}{\bibfnamefont{H.~B.} \bibnamefont{Nielsen}} \bibnamefont{and}
  \bibinfo{author}{\bibfnamefont{M.}~\bibnamefont{Ninomiya}},
  \bibinfo{journal}{Physics Letters B} \textbf{\bibinfo{volume}{130}},
  \bibinfo{pages}{389 } (\bibinfo{year}{1983}), ISSN \bibinfo{issn}{0370-2693}.

\bibitem[{\citenamefont{Aji}(2011)}]{AjiABJAnomaly}
\bibinfo{author}{\bibfnamefont{V.}~\bibnamefont{Aji}}, \bibinfo{journal}{ArXiv
  e-prints}  (\bibinfo{year}{2011}), \eprint{1108.4426}.

\bibitem[{\citenamefont{{Liu} et~al.}(2012)\citenamefont{{Liu}, {Ye}, and
  {Qi}}}]{QiWeylAnomaly}
\bibinfo{author}{\bibfnamefont{C.-X.} \bibnamefont{{Liu}}},
  \bibinfo{author}{\bibfnamefont{P.}~\bibnamefont{{Ye}}}, \bibnamefont{and}
  \bibinfo{author}{\bibfnamefont{X.-L.} \bibnamefont{{Qi}}},
  \bibinfo{journal}{ArXiv e-prints}  (\bibinfo{year}{2012}),
  \eprint{1204.6551}.

\bibitem[{\citenamefont{{Son} and {Spivak}}(2012)}]{SonSpivakWeylAnomaly}
\bibinfo{author}{\bibfnamefont{D.~T.} \bibnamefont{{Son}}} \bibnamefont{and}
  \bibinfo{author}{\bibfnamefont{B.~Z.} \bibnamefont{{Spivak}}},
  \bibinfo{journal}{ArXiv e-prints}  (\bibinfo{year}{2012}),
  \eprint{1206.1627}.

\bibitem[{\citenamefont{{Zyuzin} and {Burkov}}(2012)}]{ZyuninBurkovWeylTheta}
\bibinfo{author}{\bibfnamefont{A.~A.} \bibnamefont{{Zyuzin}}} \bibnamefont{and}
  \bibinfo{author}{\bibfnamefont{A.~A.} \bibnamefont{{Burkov}}},
  \bibinfo{journal}{ArXiv e-prints}  (\bibinfo{year}{2012}),
  \eprint{1206.1868}.

\bibitem[{\citenamefont{{Wang} and {Zhang}}(2012)}]{WeylCDW}
\bibinfo{author}{\bibfnamefont{Z.}~\bibnamefont{{Wang}}} \bibnamefont{and}
  \bibinfo{author}{\bibfnamefont{S.-C.} \bibnamefont{{Zhang}}},
  \bibinfo{journal}{ArXiv e-prints}  (\bibinfo{year}{2012}),
  \eprint{1207.5234}.

\bibitem[{\citenamefont{Hosur et~al.}(2012)\citenamefont{Hosur, Parameswaran,
  and Vishwanath}}]{HosurWeylTransport}
\bibinfo{author}{\bibfnamefont{P.}~\bibnamefont{Hosur}},
  \bibinfo{author}{\bibfnamefont{S.~A.} \bibnamefont{Parameswaran}},
  \bibnamefont{and}
  \bibinfo{author}{\bibfnamefont{A.}~\bibnamefont{Vishwanath}},
  \bibinfo{journal}{Phys. Rev. Lett.} \textbf{\bibinfo{volume}{108}},
  \bibinfo{pages}{046602} (\bibinfo{year}{2012}).

\bibitem[{\citenamefont{Burkov et~al.}(2011)\citenamefont{Burkov, Hook, and
  Balents}}]{BurkovNodalSemimetal}
\bibinfo{author}{\bibfnamefont{A.~A.} \bibnamefont{Burkov}},
  \bibinfo{author}{\bibfnamefont{M.~D.} \bibnamefont{Hook}}, \bibnamefont{and}
  \bibinfo{author}{\bibfnamefont{L.}~\bibnamefont{Balents}},
  \bibinfo{journal}{Phys. Rev. B} \textbf{\bibinfo{volume}{84}},
  \bibinfo{pages}{235126} (\bibinfo{year}{2011}).

\bibitem[{\citenamefont{Yanagishima and Maeno}(2001)}]{WeylResistivityMaeno}
\bibinfo{author}{\bibfnamefont{D.}~\bibnamefont{Yanagishima}} \bibnamefont{and}
  \bibinfo{author}{\bibfnamefont{Y.}~\bibnamefont{Maeno}},
  \bibinfo{journal}{Journal of the Physical Society of Japan}
  \textbf{\bibinfo{volume}{70}}, \bibinfo{pages}{2880} (\bibinfo{year}{2001}).

\bibitem[{\citenamefont{{Tafti} et~al.}(2011)\citenamefont{{Tafti}, {Ishikawa},
  {McCollam}, {Nakatsuji}, and {Julian}}}]{EuIridateExperiments}
\bibinfo{author}{\bibfnamefont{F.~F.} \bibnamefont{{Tafti}}},
  \bibinfo{author}{\bibfnamefont{J.~J.} \bibnamefont{{Ishikawa}}},
  \bibinfo{author}{\bibfnamefont{A.}~\bibnamefont{{McCollam}}},
  \bibinfo{author}{\bibfnamefont{S.}~\bibnamefont{{Nakatsuji}}},
  \bibnamefont{and} \bibinfo{author}{\bibfnamefont{S.~R.}
  \bibnamefont{{Julian}}}, \bibinfo{journal}{ArXiv e-prints}
  (\bibinfo{year}{2011}), \eprint{1107.2544}.

\bibitem[{\citenamefont{Schnyder et~al.}(2008)\citenamefont{Schnyder, Ryu,
  Furusaki, and Ludwig}}]{SFRLClassification}
\bibinfo{author}{\bibfnamefont{A.~P.} \bibnamefont{Schnyder}},
  \bibinfo{author}{\bibfnamefont{S.}~\bibnamefont{Ryu}},
  \bibinfo{author}{\bibfnamefont{A.}~\bibnamefont{Furusaki}}, \bibnamefont{and}
  \bibinfo{author}{\bibfnamefont{A.~W.~W.} \bibnamefont{Ludwig}},
  \bibinfo{journal}{Phys. Rev. B} \textbf{\bibinfo{volume}{78}},
  \bibinfo{pages}{195125} (\bibinfo{year}{2008}).

\bibitem[{\citenamefont{Kitaev}(2009)}]{KitaevClassification}
\bibinfo{author}{\bibfnamefont{A.}~\bibnamefont{Kitaev}},
  \bibinfo{organization}{{L.~D.~Landau Memorial Conference ``Advances in
  Theoretical Physics''}} (\bibinfo{publisher}{AIP}, \bibinfo{year}{2009}),
  vol. \bibinfo{volume}{1134}, pp. \bibinfo{pages}{22--30}.

\bibitem[{\citenamefont{Hosur et~al.}(2010)\citenamefont{Hosur, Ryu, and
  Vishwanath}}]{HosurRyuChiralTISC}
\bibinfo{author}{\bibfnamefont{P.}~\bibnamefont{Hosur}},
  \bibinfo{author}{\bibfnamefont{S.}~\bibnamefont{Ryu}}, \bibnamefont{and}
  \bibinfo{author}{\bibfnamefont{A.}~\bibnamefont{Vishwanath}},
  \bibinfo{journal}{Phys. Rev. B} \textbf{\bibinfo{volume}{81}},
  \bibinfo{pages}{045120} (\bibinfo{year}{2010}).

\end{thebibliography}

\end{document}